\documentclass{article}

\usepackage{graphicx}

\newcommand{\nn}{\nonumber\\}

\begin{document}
\baselineskip=20pt
\hsize=340pt
\vsize=490pt
\begin{titlepage}
\title{
\hfill\parbox{4cm}
{\normalsize RIKEN-TH-171
}\\
\vspace{1em}
\bf
Exploring Vacuum Structure around Identity-Based Solutions
}


\author{
Isao \textsc{Kishimoto}$^1$\footnote{E-mail address:
ikishimo@riken.jp}
\ and Tomohiko \textsc{Takahashi}$^2$\footnote{E-mail
 address: tomo@asuka.phys.nara-wu.ac.jp}\\
\\
%
$^1${\it Theoretical Physics Laboratory, RIKEN,}\\
{\it Wako 351-0198, Japan}\\
\smallskip
$^2${\it Department of Physics, Nara Women's University,}\\
{\it Nara 630-8506, Japan}
}

\date{\normalsize October, 2009}
\maketitle
\thispagestyle{empty}

\begin{abstract}
We explore the vacuum structure in bosonic open string field theory
expanded around an identity-based solution parameterized by
$a\ (\geq -1/2)$. Analyzing the expanded theory using level
truncation approximation up to level 20, we find that the
theory has the tachyon vacuum solution for $a>-1/2$.
We also find that, at $a=-1/2$, there exists an 
unstable vacuum solution in the expanded theory and
the solution is expected to be the perturbative open string vacuum.
These results reasonably support the
expectation that the identity-based solution is a trivial pure gauge
configuration for $a>-1/2$, but it can be regarded as the tachyon vacuum
solution at $a=-1/2$.
\end{abstract}
\end{titlepage}

\section{Introduction}

Bosonic open string field theory (SFT) possesses classical
solutions describing a tachyon vacuum where D-branes with attached open
strings completely annihilate. A numerical tachyon vacuum solution was
first constructed using level truncation approximation in the Siegel
gauge \cite{Sen:1999nx}. Then, an analytic classical solution has been
constructed by Schnabl \cite{Schnabl:2005gv} and it has been found that
the solution has several properties of the tachyon vacuum. The
vacuum energy of the nontrivial analytic solution exactly cancels the
D-brane tension, and the cohomology of the kinetic operator around the
vacuum is trivial \cite{Ellwood:2006ba}.

In SFT, there is another type of analytic solution that was constructed
earlier on the basis of the identity string field instead of wedge
states used in Schnabl's solution \cite{Takahashi:2002ez}.\footnote{There
are other attempts to construct analytic solutions in SFT. See, for
example, \cite{Kishimoto:2009nd} and references therein.}
Interestingly, an identity-based solution discussed in 
Refs.~\cite{Takahashi:2002ez,Kishimoto:2002xi,
Takahashi:2003ppa,Takahashi:2003xe} has some properties
of the tachyon vacuum. Moreover, for the identity-based solution, there
is a possibility of understanding the existence of closed strings on the
tachyon vacuum. 

In \S 2, we will briefly summarize
the identity-based solution. In \S 3
and \S 4, we will present new results for vacuum structure around the
identity-based solution \cite{Kishimoto:2009nd}.
Then, we propose that the solution can be
regarded as the analytic tachyon vacuum solution.

\section{Identity-based solutions in bosonic open SFT}

The identity-based solution can be expressed
as \cite{Takahashi:2002ez,Kishimoto:2002xi,Takahashi:2003ppa,Takahashi:2003xe}
\begin{eqnarray}
\label{Eq:idsol}
 \Psi_0=Q_L(e^h-1)I-C_L((\partial h)^2e^h)I,
\end{eqnarray}
where $I$ is the identity string field associated with the star product,
and the half string operators $Q_L$ and $C_L$ are defined using the
BRST current $J_B(z)$ and ghost field $c(z)$ as follows,
\begin{eqnarray}
 Q_L(f)=\int_{C_{\rm left}} \frac{dz}{2\pi i} f(z)J_B(z),\ \ \ 
 C_L(g)=\int_{C_{\rm left}} \frac{dz}{2\pi i} g(z)c(z).
\end{eqnarray}
Here, $C_{\rm left}$ denotes the contour along a right semi-unit circle
from $-i$ to $i$, which conventionally corresponds to the left half
of strings. The function $h(z)$ is defined on the whole unit circle.
For the function $h(z)$ satisfying $h(-1/z)=h(z)$ and $h(\pm i)=0$,
the equations of motion, $Q_B\Psi_0+\Psi_0*\Psi_0=0$, hold.

Hereafter, we consider the identity-based solution derived from the
function 
\begin{eqnarray}
\label{Eq:hz}
 h(z)&=&\log\left(1+\frac{a}{2}\left(z+\frac{1}{z}\right)^2\right)\\
&=& -\log(1-Z(a))^2-\sum_{n=1}^\infty
\frac{(-1)^n}{n}Z(a)^n (z^{2n}+z^{-2n}),
\end{eqnarray}
where $Z(a)=(1+a-\sqrt{1+2a})/a$.
For the solution to satisfy the reality condition,
the parameter $a$ is larger than or equal to $-1/2$.
This function gives the simplest example to capture typical feature of
the identity-based solution.

i) It has a well-defined universal Fock space expression:
if we expand the solution by Fock space states
\begin{eqnarray}
 \left|\Psi_0(a)\right>=\varphi_0(a)c_1\left|0\right>
+v_0(a)c_1 L_{-2}^X\left|0\right>
+u_0(a)c_{-1}\left|0\right>+\cdots,
\end{eqnarray}
the coefficients $\varphi_0(a)$, $v_0(a)$ and $u_0(a)$ are finite
for all $a$.

ii) For $a>-1/2$, the solution can be expressed as a pure gauge form
connecting to a trivial configuration, but
at $a=-1/2$, the solution can be given as a type of singular gauge
transformation of the trivial configuration \cite{Takahashi:2002ez}:
we can rewrite the solution for generic $a$ as
\begin{eqnarray}
 \Psi_0(a)=g(a)*Q_B g^{-1}(a),
\end{eqnarray}
but $g(a)$ becomes singular at $a=-1/2$.

These facts suggest that the
solution for $a>-1/2$ is a trivial pure gauge, but it becomes a
non-trivial configuration at $a=-1/2$.

Moreover, if we expand the string field as $\Psi=\Psi_0+\Phi$,
we can obtain the action for the fluctuation around the identity-based
solution. In the expanded theory, the kinetic operator can be 
written as
\begin{eqnarray}
\label{Eq:expkin}
 Q'&=&(1+a)Q_B+\frac{a}{2}(Q_2+Q_{-2})+4aZ(a)c_0
-2aZ(a)^2(c_2+c_{-2})\nn
&&-2a(1-Z(a)^2)\sum_{n=2}^\infty(-1)^n
Z(a)^{n-1}(c_{2n}+c_{-2n}),
\end{eqnarray}
where we expand the BRST current and ghost field as
$J_B(z)=\sum_nQ_nz^{-n-1}$ and $c(z)=\sum_n c_n z^{-n+1}$,
respectively. 

The following facts have been known for the expanded theory for
$a>-1/2$: 
\begin{enumerate}
 \item The action obtained by expanding around the solution 
       can be transformed back to the action with the
       original BRST charge \cite{Takahashi:2002ez}.
 \item The new BRST charge gives rise to the cohomology,
       which has one-to-one correspondence to the cohomology of the
       original BRST charge \cite{Kishimoto:2002xi}.
 \item The expanded theory reproduces ordinary open string
       amplitudes \cite{Takahashi:2003xe}. 
\end{enumerate}
These are consistent with the expectation that the solution
for $a>-1/2$ corresponds to a trivial pure gauge.  On the other hand,
we find completely different properties in the
expanded theory around the solution with $a=-1/2$:
\begin{enumerate}
 \item[4.] The new BRST charge has vanishing cohomology in the
	   Hilbert space with the ghost number one \cite{Kishimoto:2002xi}.
 \item[5.] The open string scattering amplitudes vanish and the
	   result is consistent with the absence of open string
	   excitations \cite{Takahashi:2003xe}. 
\end{enumerate}
From these facts, it would be reasonable to expect that the
identity-based solution at $a=-1/2$ indeed corresponds to the tachyon
vacuum solution. 
Hence, we expect that the identity-based solution corresponds
to a trivial pure gauge form for almost all the parameter region
and it can be regarded as the tachyon vacuum solution
at $a=-1/2$.

To prove our conjecture, we have to calculate the vacuum energy of the
identity-based solution directly.
Formally, the vacuum energy can be calculated as
\begin{eqnarray}
 V(\Psi_0(a))={\rm (matter\ sector)}\times {\rm (ghost\ sector)}
=\infty \times 0,
\end{eqnarray}
namely, the vacuum energy is given as an indefinite quantity.
To calculate the vacuum energy, it is necessary to apply
a kind of canonical regularization to fix the ambiguity. Indeed, the
level can be regarded as 
a regularization parameter for the numerical solution in the Siegel
gauge. Hence, the difficulty of calculating the vacuum energy seems to
arise from the lack of such a regularization method for the
identity-based solution.  

However, we can provide indirect evidence that supports the possibility
of calculating the vacuum energy. The vacuum structure in the theory
expanded around the identity-based solution has been analyzed using
level truncation approximation and then we have found the following
results: 
\begin{enumerate}
 \item[6.] A numerical analysis shows that the nonperturbative vacuum
	   found for $a>-1/2$ disappears as $a$ approaches $-1/2$
	   \cite{Takahashi:2003ppa}.
 \item[7.] The energy of the nonperturbative vacuum for $a>-1/2$
          becomes closer to the value appropriate to cancel the D-brane
	   tension as the truncation level
       increases \cite{Takahashi:2003ppa}.
\end{enumerate}
These imply that the theory around the identity-based solution
for $a>-1/2$ has the tachyon vacuum, but the theory at $a=-1/2$ is
stable.
From consistency with the theory before expanding a string field,
it follows that the vacuum energy of the identity-based solution
itself is zero for $a>-1/2$ and it is equal to the
tachyon vacuum energy at $a=-1/2$.

\section{Annihilation of tachyon vacuum
\label{sec:annihilation}}

We consider the tachyon vacuum in the expanded theory around
the identity-based solution.
The expanded theory has a gauge symmetry
under
\begin{eqnarray}
\label{eq:gauge_tr}
 \delta\Phi=Q'\Lambda+\Phi*\Lambda-\Lambda*\Phi,
\end{eqnarray}
where $Q'$ is given by Eq.~(\ref{Eq:expkin}).
To find classical solutions in the theory,
we impose the Siegel gauge condition on the fluctuation string field;
$b_0\Phi=0$.
Under the Siegel
gauge condition, the potential can be expressed as
\begin{eqnarray}
\label{Eq:expaction}
 f_a(\Phi)=2\pi^2\left(
\frac{1}{2}\left<\Phi,c_0L(a)\Phi\right>
+\frac{1}{3}\left<\Phi,\Phi*\Phi\right>\right),
\end{eqnarray}
where it is normalized as $-1$ for the tachyon vacuum solution
at $a=0$.
Here, the operator $L(a)$ is given by
\begin{eqnarray}
\label{Eq:La}
 L(a)=(1+a)L_0+\frac{a}{2}(L_2+L_{-2})+a(q_2-q_{-2})
+4(1+a-\sqrt{1+2a}),
\end{eqnarray}
where $L_n$ is a total Virasoro generator and $q_n$ is a mode of the
ghost number current.\footnote{The expression is derived from
$L(a)=\{Q',\,b_0\}$. It can be rewritten only using ghost twisted
Virasoro operators as in Ref.~\cite{Takahashi:2003xe}.} 

\begin{figure}[t]
 \begin{center}
\includegraphics[width=9cm]{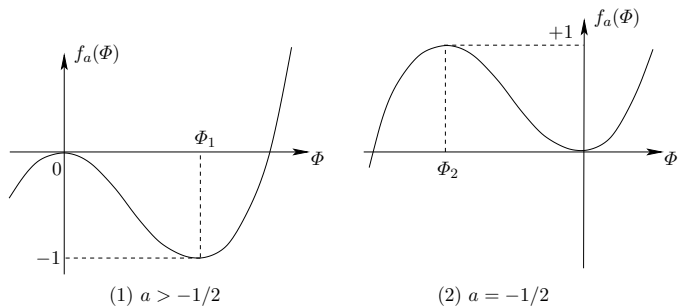}
\end{center}
\caption{Vacuum structure expected for the theory expanded around the
identity-based solution. (1) For $a>-1/2$, the theory should have a
 nontrivial vacuum solution, the vacuum energy of which cancels the
 D-brane tension. (2) At $a=-1/2$, the trivial configuration $\Phi=0$
 should be stable. 
An unstable solution is expected to exist and 
its vacuum energy should be equal to the D-brane tension.
}
\label{fig:vac}
\end{figure}

If the identity-based solution corresponds to a trivial pure gauge for
$a>-1/2$ and then to the tachyon vacuum for $a=-1/2$, the potential
(\ref{Eq:expaction}) should be illustrated as in Fig.~\ref{fig:vac}.
For $a>-1/2$, the tachyon vacuum configuration $\Phi_1$ should minimize
the potential, since the expanded theory for $\Phi$ is still
the theory on the perturbative open string vacuum. However, at $a=-1/2$, the
trivial 
configuration $\Phi=0$ should be stable since the expanded theory is
expected to be already on the tachyon vacuum. In addition, the theory at
$a=-1/2$ should have an unstable solution corresponding to the
perturbative open string vacuum of the unexpanded original theory.

\begin{figure}[t]
\begin{center}
\includegraphics[width=5cm]{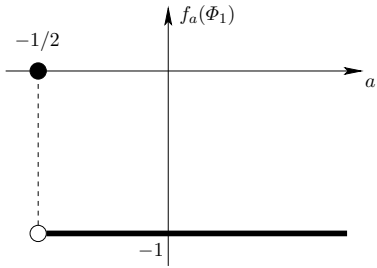}
\end{center}
\caption{Vacuum energy expected for the nontrivial stable solution
 $\Phi_1$.}
\label{fig:vacenergy1}
\end{figure}

Let us suppose that we obtain the tachyon vacuum solution $\Phi_1$ in
the expanded theory and we evaluate its vacuum energy.
Then, the vacuum energy $f_a(\Phi_1)$ is given as a function
of the parameter $a$. If the potential has such a structure as previously
expected, the vacuum energy is $-1$ for $a>-1/2$
but the solution $\Phi_1$ becomes trivial and therefore its vacuum energy
annihilates at $a=-1/2$ (see Fig.~\ref{fig:vacenergy1}).  

Now, let us consider the stable solution $\Phi_1$ by level
truncation calculation to confirm the above conjecture. Since the level
truncation is a good approximation, the vacuum energy for the
truncated solution is considered to approach the step function in
Fig.~\ref{fig:vacenergy1} as the truncation level is increased.
We apply an iterative approximation algorithm
as used in Refs.~\cite{Gaiotto:2002wy} and \cite{Kishimoto:2009cz} to
find the stable solution. 

\begin{figure}[h]
 \begin{center}
\includegraphics[width=9.5cm]{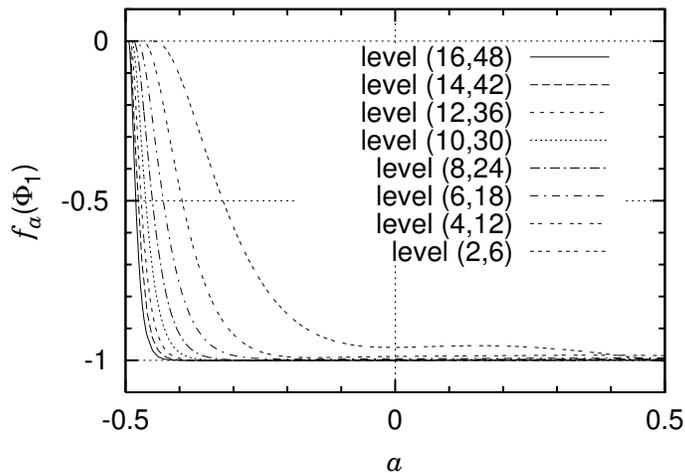}
\end{center}
\caption{Vacuum energy of the numerical stable solutions in the expanded
 theory around the identity-based solution. As the truncation level is
 increased, the resulting plots approach the step function expected
 as in Fig.~\ref{fig:vacenergy1}.} 
\label{fig:vacenergy2}
\end{figure}

We show plots of the vacuum energy for the resulting solution in
Fig.~\ref{fig:vacenergy2}. We can find that, for various $a$,
the resulting plots approach the tachyon vacuum energy $-1$
as the truncation level is increased. 
Then, for decreasing $a$ to $-1/2$, the vacuum energy increases
rapidly to zero from $-1$. As a whole, the plots become closer to the
step function 
as depicted in Fig.~\ref{fig:vacenergy1} as the truncation level
increases. These results support our expectation
for the vacuum structure associated with the stable solution $\Phi_1$.

\section{Emergence of unstable vacuum
\label{sec:emergence}}

We consider an unstable solution in the theory expanded
around the identity-based solution.
For $a>-1/2$, the expanded theory is 
unstable at $\Phi=0$. 
However, the expanded theory is expected to be
already on the tachyon vacuum for $a=-1/2$; namely, $\Phi=0$ is a stable
vacuum at $a=-1/2$. If that is the case, 
the expanded theory for $a=-1/2$ should have a nontrivial unstable
solution corresponding to the perturbative vacuum in the
original theory. 

\begin{figure}[h]
\begin{center}
\includegraphics[width=5cm]{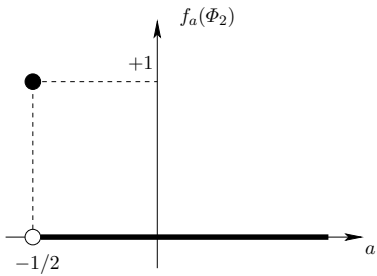}
\end{center}
\caption{Vacuum energy expected for the unstable solution
 $\Phi_2$.}
\label{fig:vacenergy4}
\end{figure}

Since the unstable solution $\Phi_2$ should correspond to the
perturbative open string vacuum, the vacuum energy of the unstable solution
should be equal to the D-brane tension (not the {\it minus} D-brane
tension). Therefore, the vacuum energy of $\Phi_2$
is expected to behave as depicted in
Fig.~\ref{fig:vacenergy4}. $f_a(\Phi_2)$ is trivially zero for $a>-1/2$,
but it should be increased to $+1$ 
owing to the emergence of the unstable solution at $a=-1/2$.

The vacuum energy and the gauge invariant overlap for $\Phi_2$ at
$a=-1/2$ are shown in Table.~\ref{tab:vacginvtab}.
We find that the vacuum energy approaches
the expected value of $+1$ as 
the truncation level is increased. 
At level $(20,60)$, the vacuum energy is about 19\% over, 
although it is about 260\% at level $(2,6)$.
Moreover, the gauge invariant overlap also takes around the expected
value of $-1$. These results suggest that the level truncation
approximation is also applicable to the analysis of the unstable
solution. Then, it is reasonably confirmed that the unstable solution
does exist as expected in the expanded theory at $a=-1/2$.

\begin{figure}[h]
 \begin{center}
\includegraphics[width=9cm]{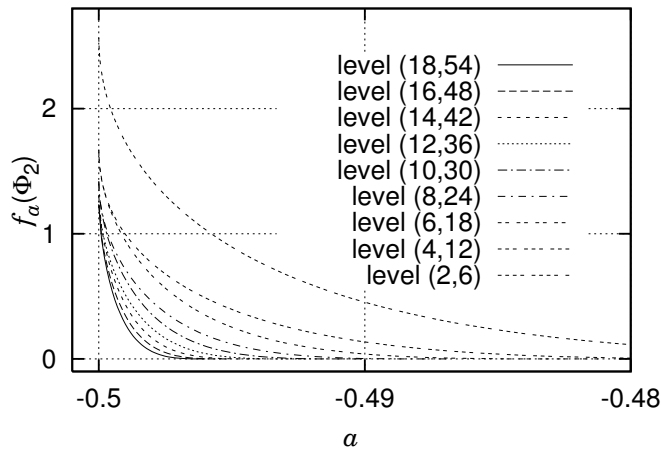}
\end{center}
\caption{Vacuum energy of the unstable solution for various $a$.}
\label{fig:unstablevac}
\end{figure}

Let us consider the unstable solution $\Phi_2$
for various $a\,(>-1/2)$.
The resulting vacuum energy of the unstable solution is depicted in
Fig.~\ref{fig:unstablevac}. The vacuum energy is around the
expected value for $a=-1/2$, but it decreases rapidly to zero for
increasing $a$.\footnote{Note that in
Fig.~\ref{fig:unstablevac} the parameter $a$ ranges from $-0.5$ to
$-0.48$ and this range is much narrower than that of
Fig.~\ref{fig:vacenergy2}.}
Thus, we find that the vacuum energy of the
unstable solution approaches the step function as expected in
Fig.~\ref{fig:vacenergy4} for increasing truncation levels.

\renewcommand{\arraystretch}{.9}
\begin{table}[h]
\caption{Vacuum energy and gauge invariant overlap for the unstable
 solution $\Phi_2$ at $a=-1/2$.}
\label{tab:vacginvtab}
\begin{center}
 \begin{tabular}{|c|c|c|}
\hline
level&\makebox[4cm]{vacuum energy} & \makebox[4cm]{gauge inv. overlap}\\
\hline
\hline
$(0,0)$ & $2.3105795$  
  & $-1.0748441$\\ 
\hline
$(2,6)$ & $2.5641847$  
  & $-1.0156983$\\ 
\hline
$(4,12)$ & $1.6550774$  
  & $-0.9539832$\\ 
\hline
$(6,18)$ &  $1.6727496$ 
  & $-0.9207572$\\ 
\hline
$(8,24)$ & $1.4193393$ 
  & $-0.9377548$\\ 
\hline
$(10,30)$ & $1.4168893$ 
   & $-0.9110994$\\ 
\hline
$(12,36)$ & $1.3035715$ 
   & $-0.9237917$\\ 
\hline
$(14,42)$ & $1.2986472$ 
   & $-0.9056729$\\ 
\hline
$(16,48)$ & $1.2357748$ 
         & $-0.9229035$\\ 
\hline
$(18,54)$ & $1.2310583$ 
         & $-0.9086563$\\ 
\hline
$(20,60)$ & $1.1915648$ 
         & $-0.9212376$\\ 
\hline
\end{tabular}\\
\end{center}
\end{table}

\section{Summary and discussion
\label{sec:summary}}

We have found that the stable and unstable solutions in the Siegel gauge,
$\Phi_1$ and $\Phi_2$, numerically exist in the theory expanded around
the identity-based solution. For these solutions, we have evaluated the
vacuum energy in terms of level truncation approximation up to level
$(20,60)$. In our paper \cite{Kishimoto:2009nd}, we have also evaluated
the gauge invariant overlap for these solutions. As the truncation level
is increased, the plots of the gauge invariants for the parameter $a$
become remarkably closer to the behavior expected from the vacuum
structure in Fig.~\ref{fig:vac}. These results strongly  support our
expectation that the identity-based solution corresponds to a trivial
pure gauge configuration for $a>-1/2$, but it can be regarded as the
tachyon vacuum solution for $a=-1/2$.  

While these results are encouraging, the direct calculation
remains one of the most difficult issues in string field theory.
However, the success of the numerical analysis is a characteristic
feature of the identity-based solution. In addition to the Schnabl-type
solutions, the identity-based solution seems to provide
complementary approaches to a deeper understanding of the string field
theory. 
In particular, we expect that the worldsheet picture in the expanded
theory clarifies the existence of closed strings on the tachyon
vacuum \cite{Drukker:2003hh,Igarashi:2005sd}.

\section*{Acknowledgements}

This work was supported in part by JSPS Grant-in-Aid for Scientific
Research (C) (\#21540269).
The work of I.~K. was supported in part by a Special Postdoctoral 
Researchers Program at RIKEN.
The work of T.~T. was supported in part by Nara Women's University
Intramural Grant for Project Research.
Numerical computations in this work were partly carried out 
on the Computer Facility of the Yukawa Institute for
Theoretical Physics in Kyoto University
and the RIKEN Integrated Cluster of Clusters (RICC) facility.

%


\end{document}